# Cloud-Magnetic Resonance Imaging System: In the Era of 6G and Artificial Intelligence

Yirong Zhou, Yanhuang Wu, Yuhan Su, Jing Li, Jianyun Cai, Yongfu You, Di Guo, Xiaobo Qu*

*Abstract*—Magnetic Resonance Imaging (MRI) plays an important role in medical diagnosis, generating petabytes of image data annually in large hospitals. This voluminous data stream requires a significant amount of network bandwidth and extensive storage infrastructure. Additionally, local data processing demands substantial manpower and hardware investments. Data isolation across different healthcare institutions hinders cross-institutional collaboration in clinics and research. In this work, we anticipate an innovative MRI system and its four generations that integrate emerging distributed cloud computing, 6G bandwidth, edge computing, federated learning, and blockchain technology. This system is called Cloud-MRI, aiming at solving the problems of MRI data storage security, transmission speed, AI algorithm maintenance, hardware upgrading, and collaborative work. The workflow commences with the transformation of k-space raw data into the standardized Imaging Society for Magnetic Resonance in Medicine Raw Data (ISMRMRD) format. Then, the data are uploaded to the cloud or edge nodes for fast image reconstruction, neural network training, and automatic analysis. Then, the outcomes are seamlessly transmitted to clinics or research institutes for diagnosis and other services. The Cloud-MRI system will save the raw imaging data, reduce the risk of data loss, facilitate inter-institutional medical collaboration, and finally improve diagnostic accuracy and work efficiency.

*Index Terms*—Magnetic Resonance Imaging, Cloud Computing, 6G Bandwidth, Artificial Intelligence, Edge Computing, Federated Learning, Blockchain.

## I. INTRODUCTION

MAGNETIC Resonance Imaging (MRI) stands as a pivotal component in medical diagnosis. Petabytes of MRI data are generated annually by healthcare facilities for diagnosis and research [1], which consume large amounts of network bandwidth, data transmission [2], and storage devices. This storage time is mandatory required up to 30 years [3, 4], and even expected for unlimited time for a personal digital life. Such a long time may let storage devices be exposed to data loss and security risk, and the lack of data backup and recovery mechanisms may lead to permanent loss of these medical data. Besides, MRI data acquired by each hospital are stored locally, limiting data sharing and cross-institutional research [5]. Differences in technology levels and resources between hospitals lead to inconsistent data processing that may affect diagnostic accuracy.

To facilitate medical data sharing, the government has introduced many policies. For example, the Chinese government has introduced a sequence of policies, such as the Guiding Opinions on Accelerating the Construction of Population Health Informatization (2013) [6], the Healthy China Action Plan (2019-2030) [7], the Opinions on Improving the Healthcare Service System (2023) [8]. Great expectation has been made to improve healthcare with medical data sharing and usage.

Even though, MRI data has not been fully utilized. Beyond the commonly saved Digital Imaging and Communications in Medicine (DICOM) format for diagnosis, MRI has its specific raw data. This data is called k-space, which is originally acquired from the imaging scanner and then reconstructed to form images, followed by post-processing and conversion into DICOM images. In the era of artificial intelligence (AI), images can be reconstructed through deep learning on local servers. However, the ongoing maintenance and updating of these algorithms, coupled with the necessary hardware, incur substantial computation power, manpower, and time. Besides, diagnosis may be more efficient if the reconstruction and conversion are skipped [9].

In this paper, we propose a Cloud-MRI system (Fig. 1), a comprehensive solution that integrates distributed cloud computing [10], 6G bandwidth [11], edge computing [12], federated learning [13], and blockchain [14]. With this system, k-space raw data can be uploaded to cloud computing servers or local edge nodes in a unified Imaging Society for Magnetic Resonance in Medicine Raw Data (ISMRMRD) [15] format via the command line, enabling advanced tasks such as the fast image reconstruction [16, 17], physics-informed synthetic data training [18], and other AI tasks. Results are distributed to cloud radiologists in each region by using a browser or mobile device

This work was supported by the National Natural Science Foundation of China (62122064, 61971361, 62331021, 62371410), the Natural Science Foundation of Fujian Province of China (2023J02005 and 2021J011184), the President Fund of Xiamen University (20720220063), and the Nanqiang Outstanding Talents Program of Xiamen University. (*Corresponding author: Xiaobo Qu, Email: quxiaobo@xmu.edu.cn)

Yirong Zhou, Yanhuang Wu, and Xiaobo Qu are with the Department of Electronic Science, Biomedical Intelligent Cloud R&D Center, Fujian Provincial Key Laboratory of Plasma and Magnetic Resonance, National Institute for Data Science in Health and Medicine, Institute of Artificial Intelligence, Xiamen University, Xiamen 361104, China.

Yuhan Su is with the Department of Electronic Science, Key Laboratory of Digital Fujian on IoT Communication, Architecture, and Safety Technology, Xiamen University, Xiamen 361104, China.

Jing Li is with Shanghai Electric Group CO., LTD, Shanghai 200001, China.
Jianyun Cai is with China Telecom Group, Quanzhou 362000, China.
Yongfu You is with China Mobile Group, Xiamen 361005, China.
Di Guo is with School of Computer and Information Engineering, Fujian Engineering Research Center for Medical Data Mining and Application, Xiamen University of Technology, Xiamen 361024, China.

for diagnosing diseases and writing reports. Radiologists will have easy access to high-quality imaging data and advanced AI support to improve global medical diagnosis and efficiency.

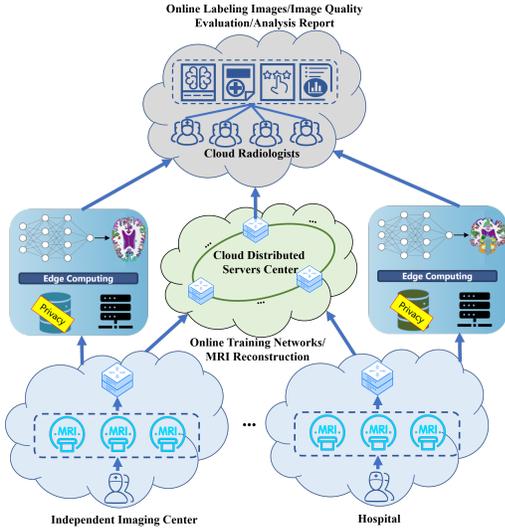

**Fig. 1.** Overview of the Cloud-MRI system.

## II. SYSTEM ARCHITECTURE

Cloud-MRI offers an end-to-end solution, starting from the raw data acquisition on a MRI scanner to form diagnostic reports written by radiologists. The architecture comprises four main parts (Fig. 2): the data transmission layer, the data processing layer and the distribution tasks layer, and a system monitoring module.

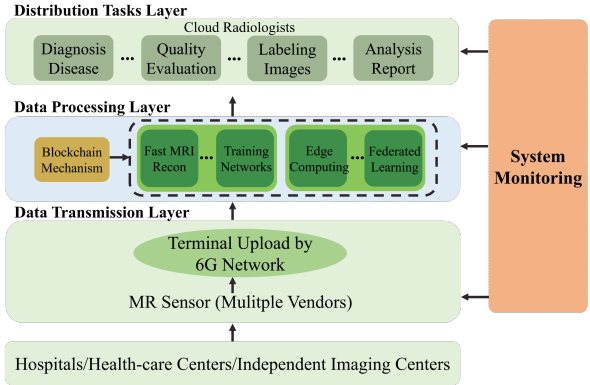

**Fig. 2.** System architecture of Cloud-MRI.

### A. Data Transmission Layer

The Cloud-MRI system facilitates seamless cross-platform and cross-vendor data exchange. The ISMRMRD format has been selected as a common storage and transmission standard. MRI raw data can be compiled into binary files with vendor identification headers, regardless of the platform or manufacturer from which the data originated.

Authorized operators initiate the upload of ISMRMRD formatted data via an encrypted channel, such as Secure Shell (SSH) by the terminal to the Cloud-MRI system. Advanced Encryption Standard (AES) [19] optimization encryption is used to ensure the privacy and integrity of the data transfer.

For fast and cost-effective transmission, the 6G network is an excellent choice. Table I highlights the remarkable synergy between 6G technology and data transmission during the scanning process. The transmission rate can be significantly improved by more than $10^4$ times and the annual fee may be saved by 35%.

TABLE I
RESOURCE COMPARISON BETWEEN LOCAL AND CLOUD

| Comparison | Local | Cloud |
| --- | --- | --- |
| Network type | Internal dedicated 4G network | Multi-region high-speed 6G network |
| File transfer rate | 100 MB/s | 1 TB/s |
| File transfer time per 10GB | 816 s | 0.01 s |
| Storage size | 64 TB | 1000 TB |
| File transfer mode | Asynchronous scan and transfer | Synchronous scan and transfer |
| Management & maintenance | Complex, specialized team management | Managed services, automated management |
| Annual fee | 1,000,000 RMB | More than 650,000 RMB |

Note: The costs of 4G and 6G are reported by China Telecom and China Mobile, respectively.

### B. Data Processing Layer

The data processing layer encompasses comprehensive components, including cloud-distributed cluster servers, local edge computing servers, federated learning techniques, and blockchain mechanisms. These components are orchestrated to support efficient processing and protection of MRI data.

Cloud-MRI system uses tools like Kubernetes and Docker to ensure the stability of cloud-distributed cluster servers. It enables state-of-the-art image reconstruction such as compressed sensing (CS) [20, 21], deep learning reconstruction [17], and physics-driven data synthesis and training [18]. For a current cloud reconstruction system [22, 23], the reconstructed image and computation time between local and cloud are compared in Fig. 3, showing that the cloud reconstruction can obtain the same high-quality images but remarkably reduce the reconstruction time.

Not limited to central computing, local edge computing servers may share a portion of the computation tasks, mitigating the cloud burden, curtailing network latency, and expeditiously catering to the needs of hospitals.

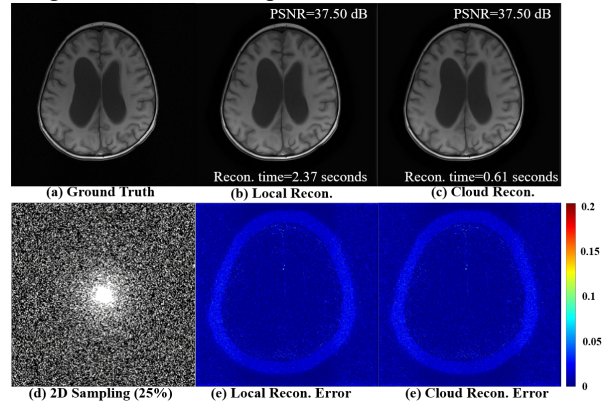

**Fig. 3.** Current deep learning reconstruction on local and cloud.

The collaborative and confidential AI training between hospitals can be conducted through federated learning tools such as PySyft [24]. It can protect data privacy since MRI raw data sharing is not necessary. In the meantime, the trainable neural network parameters can be been transmitted among hospitals, updated globally, and then sent back to edge nodes to improve the performance.

The privacy and security of the MRI data are also protected by recording the hash value of data on the blockchain and data access control by smart contracts and cryptography, which cannot be accessed by unauthorized entities.

*C. Distribution Tasks Layer*

The processing outcomes are sent back to the hospitals via an encrypted channel. Cloud radiologists can perform cloud image review, diagnostic report writing, image quality evaluation, and online labeling for medical image analysis. This versatile suite empowers cloud radiologists to access MRI data of patients from any location, facilitating remote, multidisciplinary team and cross-institute analysis. This enables patients to receive timely and accurate diagnostic and treatment services.

*D. System Monitoring*

Within the framework of the Cloud-MRI system, real-time monitoring of system operational status is diligently executed, facilitated by Security Information and Event Management (SIEM) tools [25] to ensure that the system is functioning properly and to warn of any anomalies. Meanwhile, anomaly detection through the integration of AI technology enables rapid identification of possible security threats, such as unauthorized data access or cyber-attacks.

## III. DEVELOPMENT ROUTE

The practical implementation of the Cloud-MRI system involves distinct evolutionary phases (Fig. 4). This progression unfolds across four generations, which are specialized by imaging targets and technologies, intelligent diagnosis and cross-institute cooperation modes, advanced computing, communication, and saving technologies.

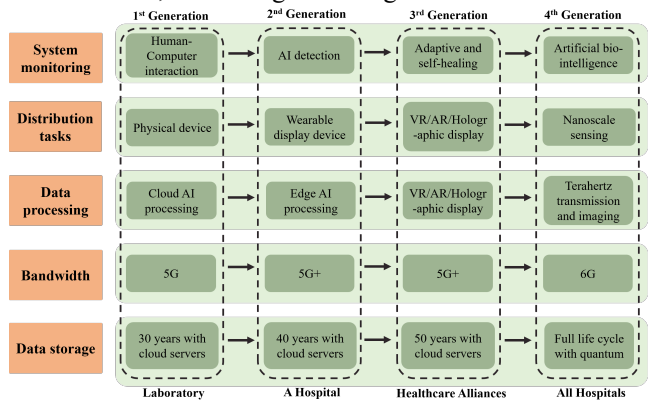

**Fig. 4.** Development route of Cloud-MRI system.

*A. First Generation*

The 1st generation architecture focuses on lab or radiology department-level infrastructure. MRI scanners mainly save the current widely used multi-contrast image information. MRI data is encrypted and transmitted to the cloud servers in ISMRMRD format for 30 years. A basic physical device interface like the web provides viewing functionality that allows radiologists to easily access and use cloud AI to process images. During this process, the system is monitored using human-computer interaction tools such as Nagios to safeguard the MRI data.

An example Cloud-MRI system (Fig. 5) has been shared at https://csrc.xmu.edu.cn/CloudBrain.html, which is based on distributed cloud computing servers. Through the web browser, k-space data can be uploaded, reconstructed, and quantified with advanced AI algorithms. Radiologists can also evaluate the image quality, and label and analyze the results without installing any software [22, 26-28]. Automatic statistical analysis and biomarker finding are also enabled [26].

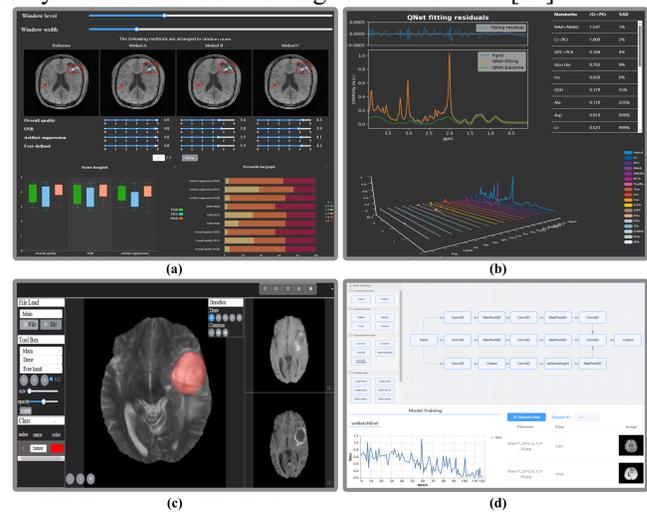

**Fig. 5.** Online cloud brain imaging system. (a) Image quality subjective evaluation and statistical analysis, (b) magnetic resonance spectroscopy quantitative analysis, (c) image labeling, (d) visualized neural network design and training. Note: "Brain" means intelligence.

*B. Second Generation*

The 2nd generation architecture will be completed in the next 3 years which aims to improve system performance within the hospitals. MRI parametric imaging is widely adopted to provide tissue property quantification [29]. Command lines (or scripts) are supported to transmit or process the data. The system also enables automatic monitoring with AI and does not require human intervention. Data storage time is increased to 40 years. To reduce the burden on cloud servers, edge AI computing will be applied. Task assignment is realized through the mobile device or wearable devices, e.g. smartwatch, to instantly view diagnostic results. The transmission latency is reduced to 0.5 milliseconds by adopting 5G+ technology.

*C. Third Generation*

The 3rd generation architecture will be completed in the next 6 years which will advance within healthcare alliances. MRI is expected to provide high-resolution multi-nuclear metabolic quantitive information [30], uncovering more diagnosis information beyond the current mostly used proton imaging. Data storage life is increased to 50 years. Virtual Reality (VR) Augmented Reality (AR), and related optical imaging devices are utilized to send diagnostic results and treatment plans to clinical doctors. Through holographic projection technology, the diagnostic results will be communicated to the radiologists, surgeons, physicians, and therapists with comprehensive information including surgical planning, survival rate, and recovery rate, and the diagnostic and treatment process will be completed after confirmation. This generation will introduce an automatic and self-healing monitoring cloud computing system and allow the bandwidth to be compatible with 5G+.

*D. Fourth Generation*

The 4th generation architecture covers all healthcare institutions with extremely powerful computing, data saving, and transmission. Very advanced MRI sensors may detect nanoscale [31] abnormal changes. These data could be saved permanently over the human full life cycle through quantum storage. If the computing ability is allowed, AI computing on atomic spins may discover the highest-resolution level of health or disease information as the MRI is firmly founded by quantum mechanics of nuclear spins. As the highlight of 6G communication, integrated sensing and transmission may also applied to MRI through terahertz communication [32], so that both MRI and terahertz imaging information may be acquired and transmitted. Comprehensive monitoring of the human lifetime is realized by using various artificial bio-intelligence such as atomic/nanoscale/Deoxyribonucleic Acid (DNA)/protein/tissue/organ-level analysis, and human inspection. The network transmission delay will be reduced to 0.1 milliseconds through 6G technology.

## IV. Conclusion

In this work, we anticipate an innovative Cloud-MRI system, that leverages an array of technologies including distributed cloud computing, 6G bandwidth, edge computing, federated learning, and blockchain. The primary objective is to address the challenges of MRI data encompassing data acquisition, storage, processing, transmission, and sharing. Additionally, the Cloud-MRI system aims to streamline the maintenance of AI algorithms, foster collaborative data initiatives across healthcare institutions, facilitate cross-institutional clinical cooperation and biomedical research, and finally improve the level and efficiency of medicine.